\documentclass[aps,prb,showpacs,reprint,superscriptaddress]{revtex4-1}
\usepackage{amsmath}
\usepackage{amssymb}
\usepackage{graphicx}
\usepackage{hyperref}
\begin{document}
\title{Highly anisotropic resistivities in the double-exchange model for strained manganites}
\author{Shuai Dong}
\affiliation{Department of Physics, Southeast University, Nanjing 211189, China}
\affiliation{Nanjing National Laboratory of Microstructures, Nanjing University, Nanjing 210093, China}
\author{Seiji Yunoki}
\affiliation{Computational Condensed Matter Physics Laboratory, RIKEN, Wako, Saitama 351-0198, Japan}
\affiliation{CREST, Japan Science and Technology Agency (JST), Kawaguchi, Saitama 332-0012, Japan}
\author{Xiaotian Zhang}
\author{Cengiz \c{S}en}
\affiliation{Department of Physics and Astronomy, University of Tennessee, Knoxville, Tennessee 37996, USA}
\affiliation{Materials Science and Technology Division, Oak Ridge National Laboratory, Oak Ridge, Tennessee 32831, USA}
\author{J.-M. Liu}
\affiliation{Nanjing National Laboratory of Microstructures, Nanjing University, Nanjing 210093, China}
\affiliation{International Center for Materials Physics, Chinese Academy of Sciences, Shenyang 110016, China}
\author{Elbio Dagotto}
\affiliation{Department of Physics and Astronomy, University of Tennessee, Knoxville, Tennessee 37996, USA}
\affiliation{Materials Science and Technology Division, Oak Ridge National Laboratory, Oak Ridge, Tennessee 32831, USA}
\date{\today}

\begin{abstract}
The highly anisotropic resistivities in strained manganites 
are theoretically studied using the two-orbital double-exchange model. At the nanoscale, the anisotropic double-exchange and Jahn-Teller distortions are found to be responsible for the robust anisotropic resistivities observed here via Monte Carlo simulations. An unbalanced in the population of orbitals caused by strain is responsible for these effects. In contrast, the anisotropic superexchange is found to be irrelevant to explain our results. Our model study suggests that highly anisotropic resistivities could be present in a wide range of strained manganites, even without (sub)micrometer-scale phase separation. In addition, our calculations also confirm the formation of anisotropic clusters in phase-separated manganites, which magnifies the anisotropic resistivities.
\end{abstract}
\maketitle

\section{Introduction}
Strongly correlated electronic materials, which are well known for the presence of complex phase competitions involving the spin, charge, and orbital degrees of freedom,\cite{Dagotto:Sci} are promising candidates to be used in new multifunctional devices.\cite{Takagi:Sci} Typically, in materials such as manganites with the colossal magnetoresistance (CMR), there are several phases with free energies that are quite close to one another but their individual physical properties can be rather different.\cite{Dagotto:Prp} Therefore, colossal responses to external perturbations, including  the CMR\cite{Tokura:Rpp} and colossal electroresistance (CER),\cite{Asamitsu:Nat} can and do occur in some manganites. During the past decade, theoretical studies on manganites have addressed many of these colossal responses, such as CMR,\cite{Burgy:Prl,Sen:Prb06,Sen:Prl,Yu:Prb} CER,\cite{Dong:Jpcm07,Dong:Prb07} surface reconstructions,\cite{Calderon:Prb,Fang:Jpsj,Dong:Apl07,Dong:Prb08} and disorder effects.\cite{Motome:Prl,Sen:Prb,Aliaga:Prb,Kumar:Prl03,Kumar:Prl,Pradhan:Prl,Kumar:Prl08,Salafranca:Prb,Chen:Prb}

In addition, the effects of strain on the properties of manganites and other complex oxides is attracting increasing attention due to the rapidly expanding research interests in complex oxides heterostructures.\cite{Dagotto:Sci07,Mannhart:Sci} In fact, phase transitions driven by strains have been discussed in manganite thin films for several years.\cite{Konishi:Jpsj,Dho:Apl,Klein:Jap,Ahn:Nat,Wu:Nm,Yamada:Apl,Dhakal:Prb,Dekker:Prb,Ding:Prl} The physical mechanism of these phase transition is mostly orbital-order-mediated.\cite{Sawada:Prb,Fang:Prl,Nanda:Prb,Dong:Prb08.3} For example, according to density functional theory (DFT) calculations, the ground states of LaMnO$_3$/SrMnO$_3$ superlattices can be tuned between A-type antiferromagnetic, ferromagnetic, and C-type antiferromagnetic phases when the ratio $c/a$ is in the range $0.96\sim1.04$, where $c$ ($a$) is the out-of-plane (in-plane) lattice constant.\cite{Nanda:Prb} Even for LaMnO$_3$ itself, the ground state may become ferromagnetic (FM)
if the $|3x^2-r^2>$/$|3y^2-r^2>$ type orbital order is fully suppressed in the cubic lattice, according to both the DFT and model calculations.\cite{Sawada:Prb,Dong:Prb08.3}

Very recently, Ward \textit{et al}. have observed high anisotropic resistivities in strained La$_{5/8-x}$Pr$_x$Ca$_{3/8}$MnO$_3$ (LPCMO) thin films.\cite{Ward:Np} LPCMO is a prototype phase-separated material.\cite{Uehara:Nat} The coexistence of FM and charge-ordered-insulating (COI) clusters at the (sub)micrometer-scale can seriously affect the electric transport properties, especially the metal-insulator transition (MIT).
The electric conductance in the phase-separated LPCMO is dominated by the percolation mechanism.\cite{Uehara:Nat,Mayr:Prl} For example, giant discrete steps in the MIT and a reemergent MIT occur in an artificially created microstructure of LPCMO when the size confinements in two directions become comparable to the phase-separated cluster sizes.\cite{Zhai:Prl,Ward:Prl} Therefore, Ward \textit{et al}. proposed that the
anisotropic percolation might be responsible for the highly anisotropic resistivities in strained LPCMO.\cite{Ward:Np} Also, our previous simulation of CER predicted anisotropic resistivities due to the electric-field-driven anisotropic percolation in phase-separated manganites.\cite{Dong:Prb07}

Then, two interesting questions arise: (1) how does strain drive the anisotropic 
percolation in the LPCMO films? And, more importantly, (2) can the large anisotropies 
occur in more standard CMR materials with nanometer-scale phase competition 
or even with bicritical clean-limit phase diagrams? Therefore, to setup a study to be used as  
a reference for future research
it is interesting to investigate theoretically with model Hamiltonians the magnitude of 
the anisotropy in transport induced by strain in cases where phase competition is present, 
but also where phase separation is not. In other words, it is important to study regimes 
where in the clean limit (no quenched disorder) a first-order transition separates the 
two competing states, typically a metal and an insulator, inducing a CMR effect in a narrow 
range of parameters, but where phase separation it not present. This calculation will allow 
us to disentangle the effects of mere strain on a clean limit model in the regime of phase 
competition from the effects of strain on a truly phase separated state. More basically, 
these investigations are important to move beyond the micrometer-scale to find the 
microscopic origin of anisotropic resistivities in generic strained manganites.

\section{Models and Techniques}
In this paper, the two-orbital double-exchange (DE) model will be employed to study the anisotropic resistivities in strained manganites. In the past decade, the DE model has been extensively studied and it proved to be a quite reasonable model to describe perovskite manganites.\cite{Dagotto:Prp} In Ward \textit{et al}.'s experiments, the anisotropic strain field splits the in-plane lattice constants along the $[100]$ and $[010]$ axes in the pseudocubic convention (or the $[101]$ and $[10\bar{1}]$ axes in the orthorhombic \textit{Pnma} convention). Thus, a modified model has to be developed to reflect the features of this strained lattice, since most previous model studies were done on cubic or square lattices.

As a well-accepted approximation for manganite models, an infinite Hund coupling is here adopted. With this useful simplification, the DE model Hamiltonian reads:
\begin{eqnarray}
\nonumber H&=&-\sum_{<ij>}^{\alpha\beta}t^{\vec{r}}_{\alpha\beta}(\Omega_{ij}c_{i\alpha}^{\dagger}c_{j\beta}+H.c.)+\sum_{<ij>}J_{\rm AF}^{\vec{r}}\vec{S}_{i}\cdot \vec{S}_{j}\\
\nonumber&&+\lambda\sum_{i}(-Q_{1i}n_{i}+Q_{2i}\tau_{xi}+Q_{3i}\tau_{zi})\\
&&+\frac{1}{2}\sum_{i}(2Q_{1i}^2+Q_{2i}^2+Q_{3i}^2).
\end{eqnarray}

In the above model Hamiltonian, the first term is the standard DE interaction. $\alpha$ and $\beta$ denote the two Mn $e_{\rm g}$-orbitals $a$ ($|x^2-y^2>$) and $b$ ($|3z^2-r^2>$). $c_{ia}$ ($c_{i\alpha}^{\dag}$) annihilates (creates) an $e_{\rm g}$ electron at orbital $\alpha$ of site $i$, with its spin parallel to the localized $t_{\rm 2g}$ spin $\vec{S}_{i}$. The nearest-neighbor (NN) hopping direction is denoted by $\vec{r}$. The Berry phase $\Omega_{ij}$ generated by the infinite Hund coupling equals $\cos(\theta_{i}/2)\cos(\theta_{j}/2)+\sin(\theta_{i}/2)\sin(\theta_{j}/2)\exp[-i(\phi_{i}-\phi_{j})]$, where $\theta$ and $\phi$ are the polar and azimuthal angles of the $t_{\rm 2g}$ spins, respectively. In strained manganites, an elongated lattice constant gives rise to more straight Mn-O-Mn bonds, thus enhancing the FM DE interaction. To mimic this effect, the in-plane DE hopping amplitudes $t^{\vec{r}}_{\alpha\beta}$
have to be set as:
\begin{eqnarray}
\nonumber t^x&=&\left(
\begin{array}{cc}
t^x_{aa} &  t^x_{ab} \\
t^x_{ba} &  t^x_{bb}
\end{array}
\right) =\frac{t_0^x}{4}\left(
\begin{array}{cc}
3 &  -\sqrt{3} \\
-\sqrt{3} &  1
\end{array}
\right)\\
t^y&=&\left(
\begin{array}{cc}
t^y_{aa} &  t^y_{ab} \\
t^y_{ba} &  t^y_{bb}
\end{array}
\right) =\frac{t_0^y}{4}\left(
\begin{array}{cc}
3 &  \sqrt{3} \\
\sqrt{3} &  1
\end{array}
\right).
\end{eqnarray}
In the rest of the manuscript,
$t_0^x$ is taken as the energy unit $t_0$ and $A_t=t_0^y/t_0^x-1$ is defined
to characterize the degree of anisotropy of the DE interaction.

The second term of the model Hamiltonian is the antiferromagnetic (AFM)
superexchange (SE) interaction between NN $t_{\rm 2g}$ spins. The SE
coefficient $J_{\rm AF}$ could also become anisotropic in the strained
lattices, which is here characterized by $A_J=J_{\rm AF}^x/J_{\rm AF}^y-1$.

The third term of the model stands for the electron-lattice coupling.
$\lambda$ is a dimensionless coefficient and $n_i$ is the $e_{\rm g}$ electronic
density at site $i$. $Q$s are phonons, including the Jahn-Teller (JT) modes
($Q_{2}$ and $Q_{3}$) and the breathing mode ($Q_{1}$):
$Q_1=(\delta_x+\delta_y+\delta_z)/\sqrt{3}$, $Q_2=(\delta_x-\delta_y)/\sqrt{2}$,
and $Q_3=(-\delta_x-\delta_y+2\delta_z)/\sqrt{6}$, where $\delta$ stands for the
length change of the oxygen coordinates in the Mn-O-Mn bonds along the axes directions. $\tau$ is the orbital
pseudospin operator, namely $\tau_{x}=c_{a}^{\dag}c_{b}+c_{b}^{\dag}c_{a}$
and $\tau_{z}=c_{a}^{\dag}c_{a}-c_{b}^{\dag}c_{b}$. The last term is the lattice elastic energy.
Note that the model used here induces cooperative distortions of the oxygen positions.

The above model Hamiltonian is numerically solved via the Monte Carlo (MC)
simulation on a two-dimensional $8\times8$ lattice. The reason for
this restriction to a two-dimensional geometry is simply practical:
simulations in three-dimensional lattices are very demanding computationally.
Thus, here $\delta_z$
is set to zero and our effort will only focus on the in-plane anisotropy.
Using standard periodic boundary conditions (PBCs),
$<Q_1>$, $<Q_2>$, and $<Q_3>$ (if $<>$ stands for averages over the whole lattice)
equals to zero. 
However, to simulate the strain effect in the JT distortion,
\emph{anisotropic} PBCs (aPBCs) should be introduced to the lattice. 
In the aPBCs for 2D lattices, $<Q_2>$ is set as a constant which can be nonzero,
while $<Q_1>$ and $<Q_3>$ remain zero. To characterize this anisotropic
JT distortion, the quantity $A_Q$ is defined as $-<Q_2>/(2\sqrt{3})$.

In Ward \textit{et al}.'s experiments, the difference between the in-plane lattice
constants is small ($\sim0.2-0.3\%$).\cite{Ward:Np} Correspondingly, the
anisotropies of interactions should be weak, implying that $A_t$, $A_J$, and
$A_Q$ must be small quantities in our study.

In our MC simulations, the average $e_{\rm g}$ density $<n>$ is chosen as $0.75$.
As discussed in previous literature,
to obtain the MIT and CMR effects, the parameters ($J_{\rm AF}$, $\lambda$) should be chosen
to be near the phase boundaries between FM and AFM COI phases.\cite{Sen:Prl,Yu:Prb}
This fine tuning of couplings could be avoided by introducing quenched disorder, but our
study will be conducted in the clean limit to setup a benchmark to decide on the origin of
strain induced transport anisotropies that are investigated experimentally.
According to the phase diagram of the two-orbital DE model for $<n>=0.75$,\cite{Dong:Prl}
the parameters $J_{\rm AF}=0.09$ and $\lambda=1.2$ are suitable and they are here
adopted as the default ones in our simulation, unless other parameters are
explicitly used. In fact, other sets of parameters near the default ones have also been
partially tested and no qualitative differences have been found. Thus, this choice
of parameters do not alter the general validation of our results and conclusions, at least
qualitatively. In the MC simulation, the first $10^{4}$ MC steps are used to reach thermal
equilibrium and another $2\times10^4$ MC steps are used for measurements.

The dc conductances, which are calculated using the Kubo formula, are in units
of $e^2/h$, where $e$ is the elementary charge and $h$ is the Planck's
constant.\cite{Verges:Cpc} The resistivities are the reciprocals of MC
averaged conductances. The normalized magnetization ($M$) is obtained from
the spin structure factor $S(\vec{k})$, at $\vec{k}=(0,0)$.\cite{Dong:Prb08}

\section{Results and Discussion}
To start the discussion of results, the original state without any anisotropic contribution
is simulated as a reference. The resistivities along both the $x$ and $y$ directions
($\rho_x$ and $\rho_y$) are calculated as a function of temperature ($T$), as shown
in Fig.~1. As expected, $\rho_x$ and $\rho_y$ are almost identical in the whole $T$
range. The small differences between $\rho_x$ and $\rho_y$ are from statistical
fluctuations during the MC simulation, and these differences should converge
to zero with increasing MC simulation times. With this set of parameters,
both $\rho_x$ and $\rho_y$ show a MIT with increasing temperature
at $T_{\rm MI}\sim0.045t_0$, which is the same approximate
location as our estimation for the Curie temperature ($T_{\rm C}$),
according to the $M-T$ curve. For a typical manganite with a MIT under
zero magnetic field, $t_0$ is roughly estimated to be in the range
$0.4-0.5$ eV.\cite{Dong:Prb08,Dong:Prb08.3} Thus, $T_{\rm MI}\sim200-260$ K in agreement with
bulk measurements.
Therefore, the set of parameters ($J_{\rm AF}=0.09$, $\lambda=1.2$) used here is suitable
to describe typical manganites, such as La$_{1-x}$Ca$_x$MnO$_3$.

\begin{figure}
\includegraphics[width=0.5\textwidth]{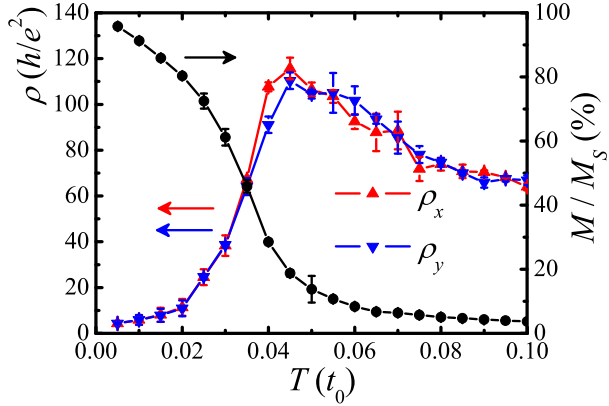}
\caption{(Color online) MC simulated resistivities (triangles)
and magnetization (dots) for a square 8$\times$8 isotropic lattice
($A_t=0$, $A_J=0$, and $A_Q=0$), as a function of $T$.}
\end{figure}

\begin{figure}
\includegraphics[width=0.5\textwidth]{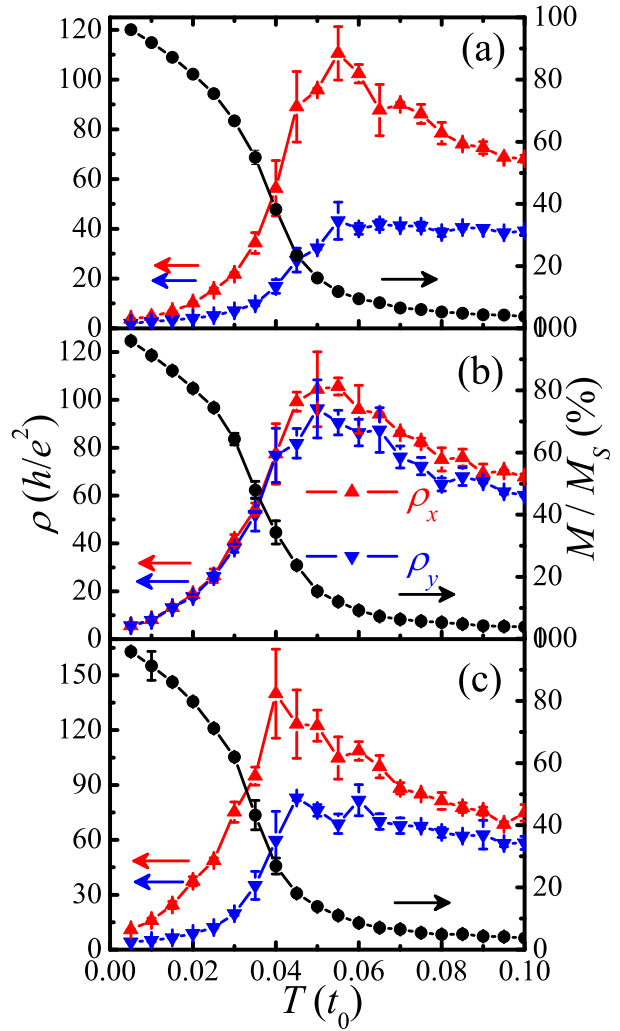}
\caption{(Color online) MC simulated resistivities (triangles) and magnetization (dots)
for strained lattices, as a function of $T$. (a) Only the anisotropic DE interaction
is here considered ($A_t=0.1$, $A_J=0$, and $A_Q=0$). (b) Only the anisotropic SE interaction
is considered ($A_t=0$, $A_J=0.1$, and $A_Q=0$). To maintain the presence of a metal-insulator
transition $J_{\rm AF}^y$ must be slightly reduced to $0.086$.
(c) Only the strained JT distortion is considered ($A_t=0$, $A_J=0$, and $A_Q=0.01$).}
\end{figure}

In the following, we will apply the aforementioned
three anisotropic interactions one by one into the model simulation
to clarify their respective roles. First, let us consider the anisotropic DE interaction. For this
purpose, $A_t$ is set to $0.1$ while other parameters are kept the same as the original ones.
In other words, the DE hopping amplitude along the $y$ direction is made $10\%$ larger
than that along the $x$ direction, because of the presence of more straight
Mn-O-Mn bonds along the $y$ axis. The resistivities and magnetization of this strained
lattice are shown in Fig.~2(a) as a function of $T$. $\rho_x$ shows a MIT similar to
the original one while $\rho_y$ is now considerably suppressed in magnitude.
Thus, a high degree of anisotropic resistivities can be obtained using an anisotropy
$A_t$ in the hoppings which is only $0.1$. Interestingly, although the difference
between $\rho_x$ and $\rho_y$ are substantial, the differences in the $T_{\rm MI}$'s
shown in $\rho_x$ and $\rho_y$ are not obvious in this $A_t=0.1$ case.
Comparing with the original one, $T_{\rm C}$ and $T_{\rm MI}$ actually simultaneously
raise to $\sim0.055t_0$  due to the increase in $t_0^y$.

Next, the anisotropic SE is taken into account. $A_t$ is restored to $0$, while
$A_J$ is set to $0.1$. In this case, $J_{\rm AF}^y$ has to be weakened slightly to preserve
the presence of an MIT, otherwise the system becomes insulating in the whole $T$ range
if $J_{\rm AF}^y$ remains at $0.09$. Thus, for this case
the new values $J_{\rm AF}^y=0.085$
and $J_{\rm AF}^x=0.0935$ are adopted. The MC simulated resistivities and magnetization
for this strained lattice are shown in Fig. 2(b), as a function of $T$.
The $T_{\rm MI}$ remains isotropic and coincides with $T_{\rm C}\sim0.05t_0$.
In contrast to the DE case, the differences between $\rho_x$ and $\rho_y$ are much smaller,
especially below $T_{\rm C}$ (or $T_{\rm MI}$): $\rho_y$ is
only slightly lower than $\rho_x$ above $0.04t_0$, and they are almost identical
below $0.04t_0$. Then we conclude that the effect of an anisotropic SE is much weaker than the
case of an anisotropic DE, when their anisotropic ratios are the same.

Finally, it is necessary to address the effect of anisotropies in the JT sector,
for completeness. In a distorted oxygen octahedron, the two $e_{\rm g}$ orbitals
are not degenerate anymore. For instance, when the lattice constants along
the $x$ and $y$ axes are different, as in the Ward \textit{et al}'s strained
manganites thin films, $<Q_2>$ is no longer zero. This nonzero $<Q_2>$ mode
induces an orbital-state ``preference'' over the whole lattice. With $A_t=0$, $A_J=0$,
and $A_Q=0.01$, the MC simulated resistivities and magnetization are shown in Fig. 2(c),
as a function of $T$. Similar to the case of an anisotropic DE, there is now a substantial
difference between $\rho_x$ and $\rho_y$. In addition, the $T_{\rm MI}$s of the
$\rho_x$ and $\rho_y$ curves becomes anisotropic: the lower resistivity curve has
a higher $T_{\rm MI}$, in agreement with the experiments.\cite{Ward:Np}.

To further clarify the anisotropic resistivities observed here,
the relative percentage difference ($\delta$) between $\rho_x$ and $\rho_y$
(defined as $\delta=(\rho_x-\rho_y)/\rho_y\times100\%$) is calculated for each of the three cases discussed above,
as shown in Fig. 3(a). For the original isotropic and the $A_J=0.1$ cases, the values of
$\delta$ are very small ($<\pm20\%$) in the whole temperature range,
as expected from Figs. 1 and 2(b).
In contrast, for the $A_t=0.1$ and $A_Q=0.01$ cases, the situation is different.
With increasing $T$ from low temperatures,
$\delta$s first increases. After each case reaches a robust peak of $200-300\%$, then they
decrease with further increases in $T$.
Interestingly, for both these two cases, the corresponding $T$s of the peaks found in $\delta$
are slightly lower than the corresponding $T_{\rm C}$s and $T_{\rm MI}$s, in agreement
with the experimental results.\cite{Ward:Np}

\begin{figure}
\includegraphics[width=0.5\textwidth]{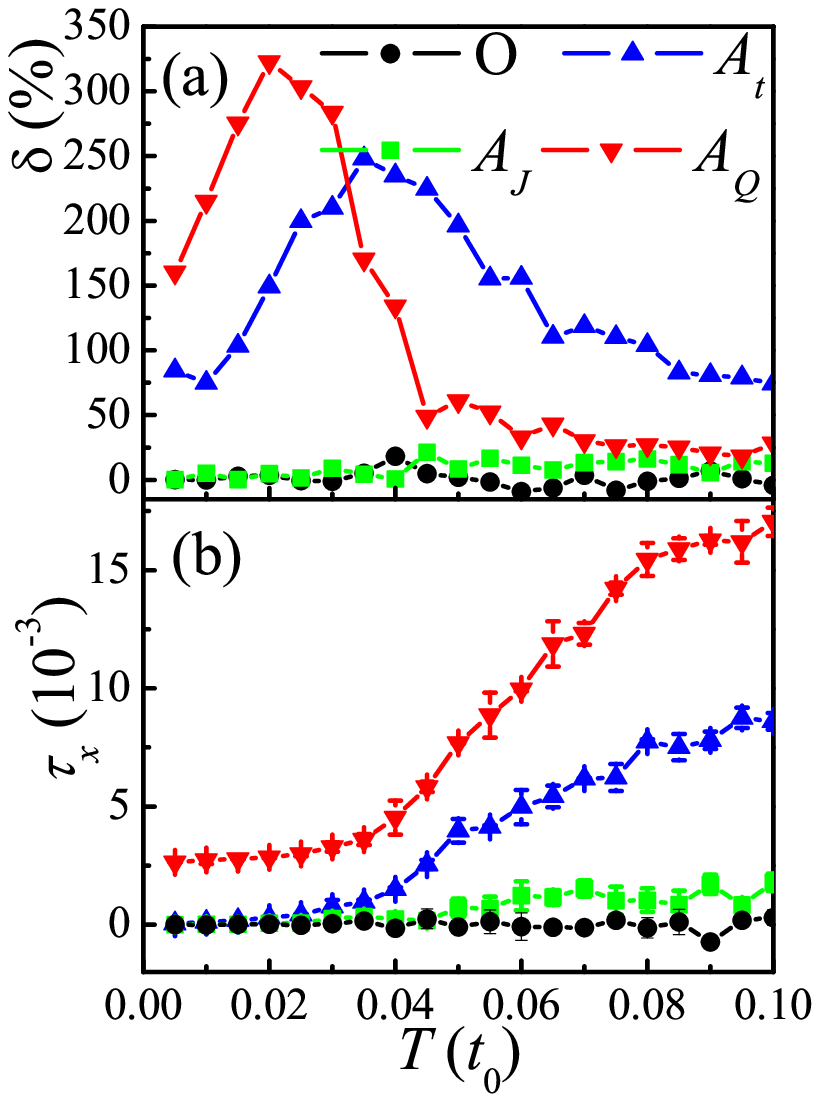}
\vskip -0.6cm
\includegraphics[width=0.4\textwidth]{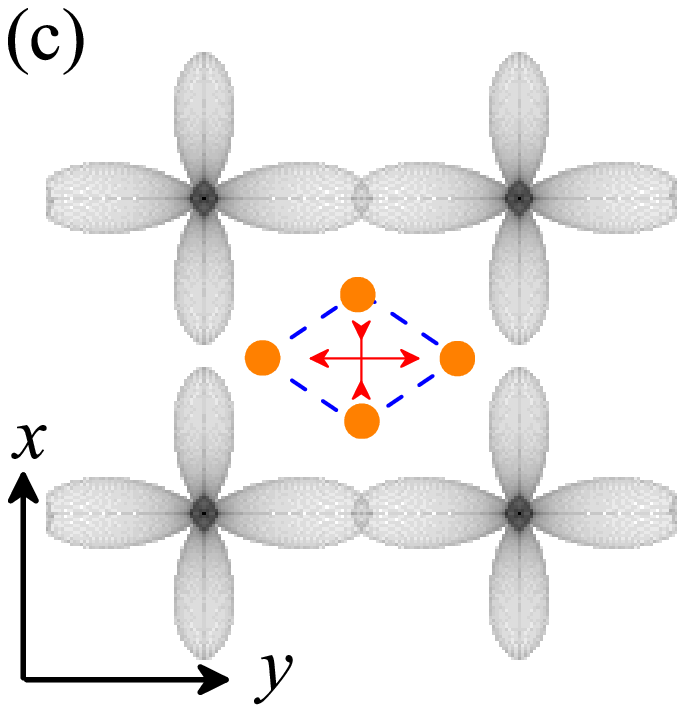}
\caption{(Color online) (a) Relative percentage differences between $\rho_x$ and $\rho_y$,
as a function of $T$, for the cases indicated. $O$ stands for original isotropic.
(b) The average values of $<\tau_x>$, as a function of $T$.
(c) Sketch of the effect of strain on the orbitals. To better distinguish the orbital-leaves along the
$x$ and $y$ directions, here $<\tau_x>/<n>$ is magnified to $0.1$. In the sketch, the overlap of
electronic clouds becomes stronger along the $y$ direction and weaker along the $x$ direction. Thus,
the conductances, which are in proportion to the overlaps, become anisotropic. Note
that the real overlaps are indirect and mediated by oxygens (not shown here). Inset: sketch of
an oxygen octahedron's in-plane distortion.}
\end{figure}

To understand the physical mechanism leading to the anisotropic resistivities
the orbital properties of the strained states, characterized by the average values of the
pseudo-spin orbital operator $<\tau_x>$, are also calculated, as shown in Fig. 3(b).
The occupation difference between the $|3y^2-r^2>$ and $|3x^2-r^2>$ components is in proportion
to $<\tau_x>$. The values of $<\tau_x>$ for the original isotropic case fluctuate around zero in
the whole $T$ range analyzed, implying that the weights of the $|3y^2-r^2>$ and $|3x^2-r^2>$
orbitals are equal, as expected by symmetry.
For the $A_J=0.1$ cases, $<\tau_x>$ still remains very small, implying that
the anisotropic $J_{\rm AF}$ used is not relevant to affect substantially the orbital composition
of the state. In fact, both these
two cases give rise to (almost) isotropic resistivities. In clear contrast, for the $A_t=0.1$
and $A_Q=0.01$ cases, finite values for $<\tau_x>$ are observed at high $T$, which are gradually
suppressed by the FM transitions with decreasing $T$. For the $A_Q=0.01$ case, the finite
$<\tau_x>$ is mainly caused by the JT distortion, which remains finite at low $T$ as long
as the lattice is anisotropically distorted. However, the finite $<\tau_x>$ for the $A_t=0.1$
case is caused by the enhanced DE process along the $y$ direction. Namely, it is a DE mediated
polarization of the orbital occupancy. Thus, for the fully FM state at low $T$, this DE mediated
orbital rearrangement is largely suppressed to near zero, which is different from the results
obtained for the JT distortion case. In summary, in our simulation the large anisotropy of
the resistivity emerges in those cases where there is an unbalanced in the orbital state population,
as sketched in Fig.~3(c), although the value of $\delta$ is not linearly dependent on $<\tau_x>$
in the whole $T$ range. In simple terms, the orbitals that increase their overlaps due to strain
are now more populated than the other ones.

Note that all the above simulations were carried out on relatively small $8\times8$ clusters
using clean-limit models
and still the anisotropy observed is comparable to that found experimentally. 
This implies that clean-limit strained manganites can be as anisotropic as phase
separated compounds.
Therefore, the LPCMO phase separation and classical percolation at the (sub)micrometer scale does
not appear to be essential to obtain highly anisotropic resistivities, but of course in
the clean limit 
the strain induced by substrates must be sufficiently large to generate a $A_t = 0.1$ 
as used here, while
for phase-separated compounds this anisotropy arises from phase competition.
Thus, the high anisotropic resistivities
should be a general properties of manganites and even other complex oxides, as long as the
bond lengths/angles are tuned to be sufficiently anisotropic by strain. 
Further experimental studies
on strained oxide films are needed to verify our results.

However, it is important to clarify that in the particular case of large scale phase-separated manganites,
the classic percolation mechanism can certainly also contribute to the anisotropic resistivities
if the shapes of the FM metallic clusters become anisotropic, as suggested in Ref.~\onlinecite{Ward:Np}.
In fact, our model can also qualitatively explain the formation of anisotropic FM clusters.
To study  an individual phase-separated FM cluster embedded in the AFM COI matrix,
the ground state energies of FM lattices with \textit{open} boundary conditions can be calculated
directly. For simplicity, all $Q_{2i}$ are set to be uniform (and equal to $<Q_2>$) and all
spins are aligned to be perfectly FM. Then, the shape of the FM clusters can be determined by varying
the lattice's shape but keeping a constant lattice area. For instance, the energies of lattices
with the same area size ($L_x\times L_y=900$, $L_x$ and $L_y$ are side lengths
along the $x$ and $y$ axes, respectively) are shown in Fig.~4(a-c), as a function of $L_x$.
The energy of the $25\times36$ lattice, which is elongated along the $y$ direction,
can be obviously more stable than that of a $30\times30$ one when $A_t>0.3$,
or $A_Q\geqslant0.04$, or $A_t\geqslant0.2$ and $A_Q\geqslant0.02$ simultaneously.
This process is qualitatively sketched in Fig.~4(d). FM clusters with other sizes
(e.g. $L_x\times L_y=576$ and $L_x\times L_y=1764$) have also been tested,
reaching the same conclusion. Thus, it is reasonable to expect similar effects when FM clusters
expand to the (sub)micrometer scale, although our microscopic model can not be directly
used on such large lattices with the currently available computational capabilities.

\begin{figure}
\includegraphics[width=0.5\textwidth]{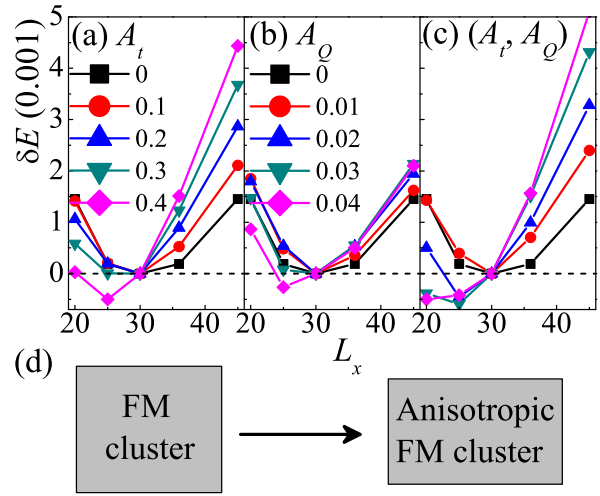}
\caption{(Color online) Energy differences of the ground states of $L_x\times L_y=900$ lattices,
as a function of $L_x$. The energy of the $L_x=30$ lattice is set as the reference point.
(a) Results obtained with the anisotropic DE interactions. (b) Results obtained with the strained
JT distortions. (c) Results obtained with both the anisotropic DE interactions and strained JT distortions
simultaneously active. The values of ($A_t$, $A_Q$) in (c) are simultaneously stepped
the same as in (a) and (b), respectively. (d) Sketch of the formation of an anisotropic
FM cluster, according to (a-c).}
\end{figure}

Finally, it is important to estimate how large should be the lattice mismatch required for the highly anisotropic resistivities observed here to appear in strained manganites with nanoscale phase separation or in the case of a bicritical phase diagram. According to the well-known Harrison's formula,\cite{Harrison:Bok} the DE hopping $t$ and SE exchange $J_{\rm AF}$ can be estimated to be in proportional to $r^{-7}$ and $r^{-14}$ ($r$ is the Mn-O-Mn's bond length), respectively. Thus $t/J_{\rm AF}$ is in proportion to $r^{7}$. To obtain the values $A_t=0.1$ and $A_J=0.1$ used in our simulations, the required lattice mismatch is about $1.4\%$. Similarly, by comparing experimental data ($r_l-r_s\sim0.6$ \AA, where $r_l$ and $r_s$ are long and short bonds, respectively)\cite{Alonso:Ic} and theoretical parameters ($\lambda|Q_2|=1.5$)\cite{Dong:Prb08.2} for the JT distortions in $R$MnO$_3$, the parameter $A_Q=0.01$ used here is estimated as $\sim0.45\%$. It should be noted that the required strain (lattice mismatch in real films, or ($A_t$, $A_Q$) in our simulations) depends on the particular materials under study (or, equivalently, the actual values of the parameters ($J_{\rm AF}$, $\lambda$) in our simulations). The anisotropies are more sensitive to strain when the system moves closer to the phase boundary between the FM and COI phases. With this idea in mind, it is natural that the anisotropies of LPCMO films can be notorious even if the lattice mismatch is small in average ($0.2-0.3\%$ in Ward \textit{et al}.'s experiments), because LPCMO is precisely at the FM-COI phase boundary. According to our simulations, the highly anisotropic resistivities are also expected in other strained CMR manganites films (even without phase separation), although the required strain might be somewhat larger than for the LPCMO case.

\section{Conclusions}

In conclusion, the high anisotropic resistivities of strained manganites films were studied
using microscopic models. For this purpose, the two-orbital double-exchange model was modified
to include the strain contributions. In this revised model, the anisotropic Jahn-Teller distortion
was emphasized, in addition to the anisotropic exchanges. The results of our MC simulation shows
that the highly anisotropic resistivities are associated with an unbalanced in orbital populations
which is driven by the anisotropic double-exchange and anisotropic Jahn-Teller distortions.
In contrast, the anisotropic superexchange was not found to be a dominant driving force for the
anisotropic resistivities. The observed high anisotropic resistivities in our simulation did not
rely on phase separation at the (sub)microscopic scale. Therefore, it is expected that this
anisotropic state could be realized in a variety of manganites and other complex oxides as well,
if a sufficiently large lattice mismatch can be achieved in the growth of the manganite films.
In addition, for the particular case of phase-separated manganites, our model investigations suggest that
the anisotropic double-exchange and strained Jahn-Teller distortions could indeed reshape the
ferromagnetic clusters, thus inducing an anisotropic percolation and concomitant
anisotropic resistivity that further enhances these effects.

\vspace{0.5cm}
\textbf{~~~~~~~~~~~ACKNOWLEDGMENTS}
\vspace{0.5cm}

We thank T.Z. Ward and J. Shen for fruitful discussions. Work was supported by the 973 Projects
of China (2006CB921802, 2009CB623303) and the National Science Foundation of China (50832002). S.Y. was
supported by CREST-JST. X.T.Z, C.S., and E.D. were supported by the USA National Science Foundation grant DMR-0706020 and by the Division of Materials Science and Engineering, Office of Basic Energy Sciences,
U.S. Department of Energy.

\bibliographystyle{apsrev4-1}
\bibliography{../ref}
\end{document}